\documentclass[tradiabstract]{aa} 

\usepackage{graphicx}
\usepackage{txfonts}

\usepackage{epsf}
\usepackage{rotating}
\usepackage{graphics}

\def \sophie{{\it SOPHIE}}
\def \MJ{M$_{\mathrm{Jup}}$}

\def \kms{km\,s$^{-1}$\/}
\def \ms{m\,s$^{-1}$\/}
\def \1s{$1\,\sigma$}
\def \kid{$\chi^2$}
\def \t0{T$_0$}

\def \jk07{Johns-Krull et al.~(\cite{jk07})}

\begin{document}

   \title{Misaligned spin-orbit in the XO-3 planetary system?\thanks{Based 
            on observations collected with the \sophie\ spectrograph on the 1.93-m telescope at
            Observatoire de Haute-Provence (CNRS), France, by the \sophie\ Consortium 
            (program 07A.PNP.CONS).}}
      
   \author{G.~H\'ebrard\inst{1}, 
                 F.~Bouchy\inst{1}, 
                 F.~Pont\inst{2}, 
                 B.~Loeillet\inst{1,3}, 
                 M.~Rabus\inst{4}, 
                 X.~Bonfils\inst{5,6}, 
                 C.~Moutou\inst{3}, 
                 I.~Boisse\inst{1},
                 X.~Delfosse\inst{6},
                 M.~Desort\inst{6},
                 A.~Eggenberger\inst{6},  
                 D.~Ehrenreich\inst{6},
                 T.~Forveille\inst{6}, 
                 A.-M.~Lagrange\inst{6},
                 C.~Lovis\inst{7},
                 M.~Mayor\inst{7},
                 F.~Pepe\inst{7},
                 C.~Perrier\inst{6}, 
                 D.~Queloz\inst{7},
                 N.~C.~Santos\inst{5,7},
                 D.~S\'egransan\inst{7},
                 S.~Udry\inst{7}, 
                 A.~Vidal-Madjar\inst{1}}

   \institute{
Institut d'Astrophysique de Paris, UMR7095 CNRS, Universit\'e Pierre \& Marie Curie, 
98bis boulevard Arago, 75014 Paris, France 
\and
Physikalisches Institut, University of Bern, Sidlerstrasse 5, 3012 Bern, Switzerland
\and
Laboratoire d'Astrophysique de Marseille, Universit\'e de Provence, CNRS (UMR 6110), 
BP 8, 13376 Marseille Cedex 12, France
\and
Instituto de Astrof{\'\i}sica de Canarias, La Laguna, Tenerife, Spain
\and
Centro de Astrof{\'\i}sica, Universidade do Porto, Rua das Estrelas, 4150-762 Porto, Portugal
\and
Laboratoire d'Astrophysique de Grenoble, 
CNRS (UMR 5571), Universit\'e J. Fourier, BP53, 38041 Grenoble, France
\and
Observatoire de Gen\`eve,  Universit\'e de Gen\`eve, 51 Chemin des Maillettes, 1290 Sauverny, Switzerland
}

   \date{Received TBC; accepted TBC}
      
  \abstract{
    The transiting extrasolar planet XO-3b is  remarkable, with a high mass and eccentric orbit. These
    unusual characteristics 
   make it interesting to test whether its orbital plane is parallel to the equator of its host star, as it 
   is observed for other transiting planets.
  We performed radial velocity measurements of XO-3 with the \sophie\ spectrograph at the 1.93-m 
   telescope of Haute-Provence Observatory during a planetary transit, and at other orbital phases. 
   This allowed us to observe the Rossiter-McLaughlin effect and, 
   together with a new analysis of the transit light curve, to refine the parameters of the planet.
  The unusual shape of the radial velocity anomaly during the transit provides a hint for a 
   nearly transverse Rossiter-McLaughlin effect. The sky-projected angle between the planetary 
   orbital axis and the stellar rotation axis should be  $\lambda =70^{\circ}\pm15^{\circ}$ 
   to be compatible with our observations.    
   This suggests that some close-in planets might result from gravitational interaction between planets 
   and/or stars rather than migration due to interaction with the accretion disk. 
   This surprising result requires confirmation by additional observations, especially 
   at lower airmass, to fully exclude the possibility that the signal is due to systematic effects. }


  \authorrunning{H\'ebrard et al.}

   \maketitle


\section{Introduction}
\label{sect_intro}

Johns-Krull et al.~(\cite{jk07}) announced the detection of \object{XO-3b}, an extra-solar planet 
transiting its F5V parent star with a 3.2-day orbital period. 
Transiting planets are of particular interest as they allow 
measurements of parameters including orbital inclination and planet radius, mass and density. 
Moreover, follow-up observations can also be performed during transits or anti-transits, yielding
physical constraints on planetary atmospheres. 

Among the forty transiting extra-solar planets known to date, XO-3b is particular 
as it is among the few on an eccentric orbit, together with HD\,147506b 
(Bakos et al.~\cite{bakos07}), HD\,17156b (Fischer et al.~\cite{fischer07}; 
Barbieri et al.~\cite{barbieri07}), 
and GJ\,436b (Butler et al.~\cite{butler04}; Gillon et al.~\cite{gillon07}). 
XO-3b is also the most massive transiting planet known to date. 
Most of the sixty known extrasolar planets, with and without transits, with orbital periods 
shorter than five days 
have masses below 2\,\MJ; XO-3b is actually one of the rare massive close-in
planets. It is just at the limit between low-mass brown dwarfs and massive 
planets, 13\,\MJ, 
which is defined by the deuterium burning limit. 
There was a quite large uncertainty on the planetary parameters of XO-3b 
and its host star. Indeed, Johns-Krull et al.~(\cite{jk07}) presented a spectroscopic 
analysis favoring large masses and radii 
($M_\mathrm{p} \simeq 13.25$~\MJ, $R_\mathrm{p} \simeq 1.95$~R$_\mathrm{Jup}$, 
$M_\star \simeq 1.41$~M$_\odot$, and $R_\star \simeq 2.13$~R$_\odot$),
whereas their light curve analysis suggests lower values 
($M_\mathrm{p} \simeq 12.03$~\MJ, $R_\mathrm{p} \simeq 1.25$~R$_\mathrm{Jup}$, 
$M_\star \simeq 1.24$~M$_\odot$, and $R_\star \simeq 1.48$~R$_\odot$)
[see however Sect.~\ref{sect_winn} and Winn et al.~(\cite{winn08a})]. 

The fast rotating star XO-3 ($V \sin I = 18.5$~\kms; Johns-Krull et al.~\cite{jk07}) 
is a favorable object for Rossiter-McLaughlin effect 
observations. This effect (Rossiter~\cite{rossiter24}; McLaughlin~\cite{mclaughlin24}) 
occurs when an object transits in front of a rotating star, causing a distortion of the 
stellar lines profile, and thus an apparent anomaly in the measured radial velocity 
of the star. The shape of the disturbed radial velocity curve allows one to determine whether 
the planet is orbiting in the same direction as its host star is rotating, and more generally 
to measure the sky-projected angle between the planetary orbital axis and the stellar 
rotation axis, usually noted $\lambda$ (see, e.g., Ohta et al.~\cite{ohta05}; 
Gim\'enez~\cite{gimenez06a}; Gaudi \& Winn~\cite{gaudi07}). 
A stellar spin axis not aligned with the orbital angular momentum 
of a planet ($\lambda \neq 0^{\circ}$) 
could reflect processes in the planet formation and 
migration, or interactions with perturbing~bodies 
(see, e.g., Malmberg et al.~\cite{malmberg07}, 
Chatterjee et al.~\cite{chatterjee07}, Nagasawa 
et al.~\cite{nagasawa08}). Solar System asteroids 
are examples of objects whose orbital axes can be misaligned from 
the Sun spin axis by over $30^{\circ}$.

Up to now, spectroscopic transits have been detected for eight exoplanets:
HD\,209458b (Queloz et al.~\cite{queloz00}), HD\,189733b (Winn et al.~\cite{winn06}), 
HD\,149026b (Wolf et al.~\cite{wolf07}), TrES-1 (Narita et al.~\cite{narita07}), 
HD\,147506b (Winn et al.~\cite{winn07}; Loeillet et al. \cite{loeillet07}), 
HD\,17156b (Narita et al.~\cite{narita08}), CoRoT-Exo-2b (Bouchy et al.~\cite{bouchy08})
and TrES-2  (Winn et al.~\cite{winn08b}). For all of these targets the stellar rotation is prograde 
relative to the planet orbit, and the sky-projected $\lambda$ angle is close to zero for most 
of them. So the axes of the stellar spins are probably parallel to the orbital axes, 
as expected for planets that formed in a protoplanetary disc far from the star and that later 
migrated closer-in. Three systems have error bars on the $\lambda$ angle that 
do not include $0^{\circ}$: 
TrES-1 ($\lambda=30^{\circ}\pm21^{\circ}$), 
CoRoT-Exo-2b ($\lambda=7.2\pm4.5^{\circ}$) and 
HD\,17156b ($\lambda=62^{\circ}\pm25^{\circ}$). 
However, those cases have the largest error bars on $\lambda$, 
and no firm detection of misalignment has yet been claimed.
Barbieri et al.~(\cite{barbieri08}) recently presented new radial velocity measurements 
of HD\,17156 secured during a transit, which agree with a spin-orbit alignment.

Approximate spin-orbit alignment therefore seems typical for exoplanets, as it is for 
planets in the Solar System. The unusual parameters of XO-3b make a test of whether 
it agrees with this apparent behavior interesting. We present here new measurements 
of XO-3 radial velocity performed during a transit and at other orbital phases. These data 
refine the orbital parameters and provide a hint of detection for a transverse 
Rossiter-McLaughlin effect, {\it i.e.}\ Êa $\lambda$ angle possibly near $90^{\circ}$. 
We also present a revised analysis of the transit light curve.

\section{Observations}
\label{sect_obs}

We observed the host star XO-3 (GSC\,03727-01064, $m_V=9.91$) with the \sophie\   
instrument at the 1.93-m telescope of Haute-Provence Observatory, France. 
\sophie\ is a cross-dispersed, environmentally stabilized echelle spectrograph dedicated to 
high-precision radial velocity measurements (Bouchy et al.~\cite{bouchy06}).   
We used the high-resolution mode (resolution power $R=75,000$) of the 
spectrograph, and the fast-read-out-time mode of the $4096\times2048$ 
15-$\mu$m-pixel CCD detector.
The two  optical-fiber circular apertures were used; the first one was 
centered on the target, and the second one was on the sky to simultaneously 
measure its background. 
This second aperture, 2' away from the first one, was used to estimate the 
spectral pollution due to the moonlight, which can be quite 
significant in these 3''-wide 
apertures (see Sect. \ref{sect_Data_reduction}).

We acquired 36 spectra of XO-3 during the night of January 28th, 2008  
(barycentric Julian date BJD$\;= 2\,454\,494.5$),  where a~full coverage 
of the planetary transit was observed. Another~19~spectra 
were acquired at other orbital phases during the following two months. 
Table~\ref{table_rv} summaries the 55 spectra finally~acquired.

\begin{table}[h]
  \centering 
  \caption{Radial velocities of XO-3 measured with \sophie.}
  \label{table_rv}
\begin{tabular}{ccccc}
\hline
\hline
BJD & RV & $\pm$\1s & exp. time & S/N p. pix. \\
-2\,400\,000 & (\kms) & (\kms) & (sec) &  (at 550 nm)  \\
\hline
\multicolumn{3}{l}{{\hspace{-0.2cm}}Planetary transit:} \\
54494.4461 & -11.298 & 0.020 & 600 & 54 \\
54494.4526 & -11.300 & 0.026 & 403 & 42 \\
54494.4578 & -11.345 & 0.027 & 373 & 40 \\
54494.4625 & -11.379 & 0.027 & 370 & 40 \\
54494.4675 & -11.390 & 0.028 & 370 & 39 \\
54494.4721 & -11.354 & 0.028 & 370 & 39 \\
54494.4767 & -11.460 & 0.029 & 370 & 38 \\
54494.4813 & -11.448 & 0.028 & 381 & 39 \\
54494.4861 & -11.419 & 0.027 & 380 & 40 \\
54494.4913 & -11.411 & 0.027 & 380 & 40 \\
54494.4960 & -11.457 & 0.028 & 380 & 39 \\
54494.5007 & -11.463 & 0.028 & 380 & 39 \\
54494.5054 & -11.549 & 0.028 & 380 & 39 \\
54494.5101 & -11.482 & 0.028 & 380 & 40 \\
54494.5152 & -11.589 & 0.028 & 380 & 39 \\
54494.5202 & -11.652 & 0.028 & 434 & 39 \\
54494.5254 & -11.611 & 0.027 & 403 & 40 \\
54494.5303 & -11.679 & 0.027 & 392 & 40 \\
54494.5352 & -11.691 & 0.036 & 395 & 40 \\
54494.5413 & -11.667 & 0.035 & 509 & 43 \\
54494.5474 & -11.782 & 0.034 & 500 & 44 \\
54494.5535 & -11.700 & 0.034 & 500 & 42 \\
54494.5599 & -11.649 & 0.035 & 554 & 44 \\
54494.5665 & -11.785 & 0.034 & 531 & 44 \\
54494.5738 & -11.811 & 0.036 & 530 & 44 \\
54494.5806 & -11.783 & 0.033 & 602 & 45 \\
54494.5880 & -11.867 & 0.033 & 622 & 44 \\
54494.5957 & -11.767 & 0.033 & 653 & 45 \\
54494.6036 & -11.785 & 0.035 & 650 & 46 \\
54494.6123 & -11.740 & 0.034 & 682 & 45 \\
54494.6210 & -11.783 & 0.035 & 775 & 45 \\
54494.6305 & -11.880 & 0.033 & 801 & 45 \\
54494.6407 & -11.843 & 0.034 & 906 & 44 \\
54494.6532 & -11.920 & 0.036 & 964 & 42 \\
54494.6668 & -11.964 & 0.037 &1321& 43 \\
54494.6822 & -11.913 & 0.049 &1300& 38 \\
\hline
\multicolumn{3}{l}{{\hspace{-0.2cm}}Other orbital phases:} \\
54496.2649 & -12.723 & 0.050 &1202 & 22 \\
54497.2609 & -10.156 & 0.029 & 655  & 37 \\
54499.2765 & -13.006 & 0.030 &1003 & 36 \\
54501.2926 & -12.433 & 0.031 & 775  & 34 \\
54501.4628 & -12.756 & 0.033 & 775  & 34 \\
54502.2730 & -13.068 & 0.024 & 614  & 44 \\
54503.2614 & -10.936 & 0.030 & 907  & 35 \\
54503.4700 & -10.182 & 0.040 &1806 & 29 \\
54504.4321 & -12.398 & 0.028 & 645  & 39 \\
54505.2889 & -13.132 & 0.023 & 635  & 46 \\
54506.2904 & -11.593 & 0.025 & 755  & 43 \\
54511.4534 & -13.041 & 0.038 & 600  & 48 \\
54512.4618 & -12.246 & 0.036 & 600  & 41 \\
54513.3091 & -10.360 & 0.052 & 600  & 48 \\
54516.3517 & -10.176 & 0.046 & 600  & 39 \\
54516.4540 & -10.267 & 0.071 & 802  & 32 \\
54551.3044 & -10.316 & 0.032 &1274 & 37 \\
54553.3002 & -13.135 & 0.033 & 999  & 36 \\
54554.3114 & -11.004 & 0.019 & 999  & 57 \\
\hline
\end{tabular}
\end{table}

The exposure times range from 6 to 30 minutes in order to reach as constant the signal-to-noise 
ratio as possible. Indeed, \sophie\ radial velocity measurements are currently affected 
by a systematic effect at low signal-to-noise ratio, which is probably due to CCD charge transfer 
inefficiency that increases at low flux level. A constant signal-to-noise ratio through a sequence
of observations reduces this uncertainty. The different exposure times needed to reach similar 
signal-to-noise ratios reflect the variable throughputs obtained, due to various atmospheric 
conditions (seeing, thin clouds, atmospheric dispersion). The sky was clear on the night whom 
the XO-3b transit was observed, but the airmass ranged from 1.2 to 3.1 during this $\sim$6-hour 
observation sequence; the exposure times therefore increased during the transit observation. 
They remain short enough to provide a good time sampling (20 measurements during the 
$\sim$3 hours of the transit). The 19 measurements outside the transit night were performed 
at airmasses  better than 1.4 but with conditions varying from photometric to cloudy.

Exposures of a thorium-argon lamp were performed every 2-3 hours during each 
observing night. Over 2-3 hours, the observed drifts were typically $\sim3$~\ms, which is thus the 
accuracy of the wavelength calibration of our XO-3 \sophie\ spectra; this is good enough for 
the expected signal. We did not use simultaneous calibration to keep the second aperture 
available for sky background estimation. During the night of January 28th, 2008, we performed 
a thorium-argon exposure before the transit and another one after the sequence, about six 
hours later. The measured drift was particularly low this night, 
1~\ms\  in six hours, which makes us confident 
the wavelength calibration did not unexpectedly drift during the observation of 
the XO-3b transit.

\section{Data reduction}
\label{sect_Data_reduction}

We extracted the spectra from the detector images and measured the radial 
velocities using the \sophie\ pipeline. 
Following the techniques described by Baranne et al.~(\cite{baranne96}) and  
Pepe et al.~(\cite{pepe02}), the radial velocities were obtained from a weighted 
cross-correlation of the spectra with a numerical mask. 
We used a standard G2 mask constructed from the Sun spectrum 
atlas including more than 3500 lines, which is well adapted to the F5V star XO-3. 
We eliminated the first eight spectral orders of the 39 available ones from 
the cross-correlation;  these blue orders are particularly noisy, especially for 
the spectra obtained at the end of the transit, when the airmass was high.
The resulting cross-correlation functions (CCFs) were fitted by Gaussians to get the 
radial velocities, as well as the width of the CCFs and their contrast with respect to  
the continuum. The uncertainty on the radial velocity was computed from the 
width and contrast of the CCF and the signal-to-noise ratio, using 
the empirical relation detailed by Bouchy et al.~(\cite{bouchy05}) and Collier Cameron et 
al.~(\cite{cameron07}). It was typically around 25~\ms\ during the night of the transit, and 
between 20 and 45~\ms\  the remaining nights. The large $V \sin I$ of this rotating stars 
makes the uncertainty slightly larger than what is usually obtained 
for such signal-to-noise ratios with~\sophie.

Some measurements were contaminated by the sky background, including mainly the 
moonlight. As the G2 mask matches the XO-3 spectrum as well as the Sun spectrum 
reflected by the Moon and the Earth atmosphere, the moonlight contamination can 
distort the shape of the CCF and thus shift the measured radial velocity. 
During our observations, the 29-\kms\ wide (FWHM) CCF of XO-3 is at radial velocities 
between $-13$ and $-10$~\kms, whereas the moonlight was centered near the 
barycentric Earth radial velocity, between $-23$ and $-20$~\kms. Thus moonlight 
contamination tends to blueshift the measured radial velocities. Following the method 
described in Pollacco et al.~(\cite{pollacco08}) and Barge et al.~(\cite{barge08}), 
we estimated the Moon contamination thanks to the second aperture, targeted on the sky, 
and then subtracted the sky CCF from the star CCF (after scaling by the throughput of the two 
fibers). Five exposures with too strong contamination were not used.
We estimated the accuracy of this correction on one hand by correcting uncontaminated 
spectra, and on the other hand by correcting uncontaminated spectra on which we have 
added moonlight contaminations. Comparisons of the corrected velocities to the uncontaminated
ones show that the method works well up to $\sim500$~\ms\ shifts, with an uncertainty of 
1/9 of the correction to which a minimum uncertainty of 25~\ms\ is quadratically added. 

The second half of the transit night measurements was contaminated by moonlight, with sky 
CCFs contrasted between 2 and 5\,\% of the continuum, whereas the XO-3 CCF has a 
contrast of 8\,\%. This implied sky corrections $<150$~\ms\ (except for the very last 
exposure where it was $\sim300$~\ms), with uncertainties in the range 25-30~\ms\ (40~\ms\  
for the last exposure). Five exposures obtained later at different phases were contaminated 
by the moonlight; corrections of 100 to 500~\ms\ were computed, with uncertainties 
in the range 30-60~\ms.

\begin{figure}[h] 
\begin{center}
\includegraphics[scale=0.465]{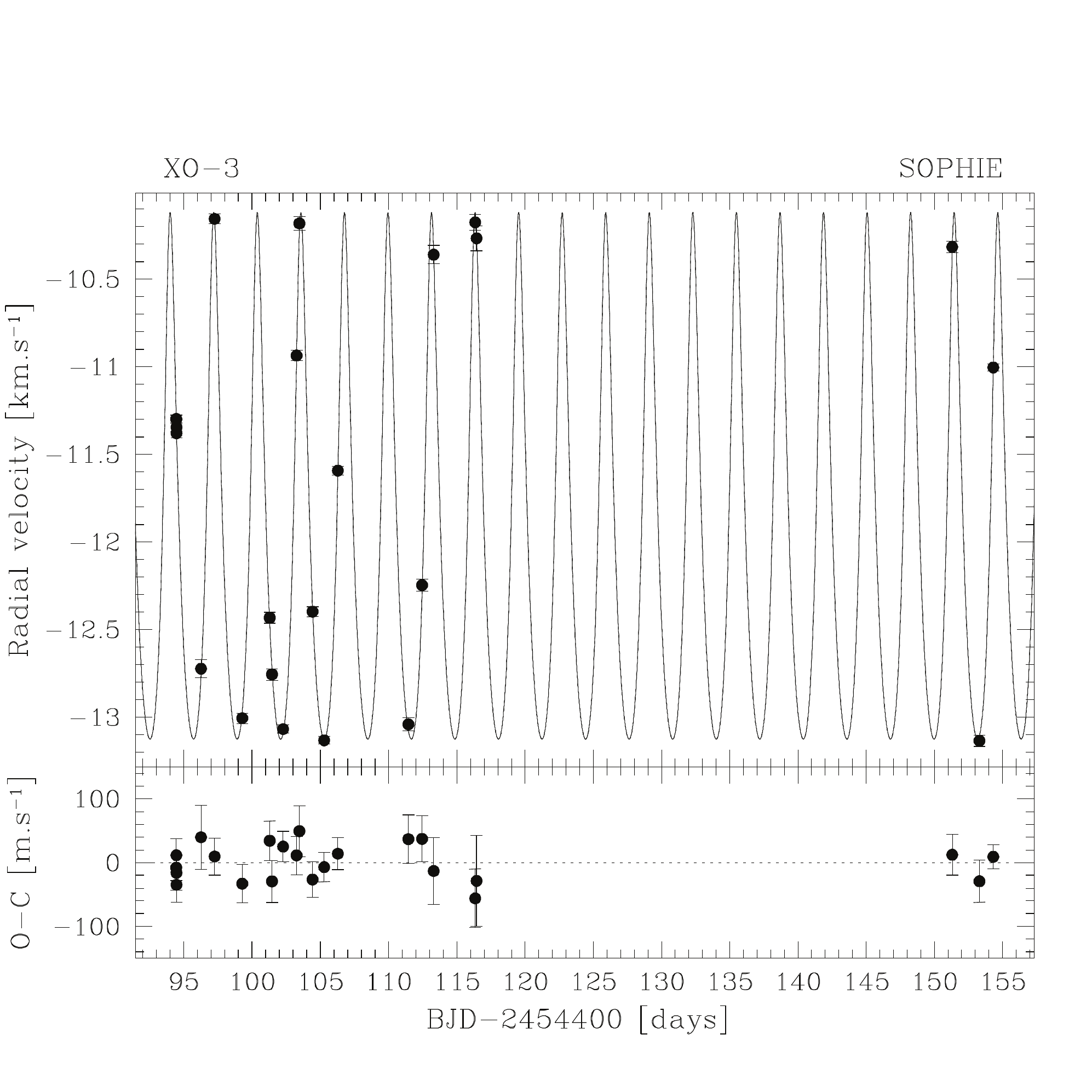}
\caption{\textit{Top:} Radial velocity measurements of XO-3 as a function of time, 
and Keplerian fit to the data (without transit). Only the 23 measurements used for
the fit are displayed. The orbital parameters corresponding to this 
fit are reported in Table~\ref{table_parameters}. 
\textit{Bottom:} Residuals of the fit with 1-$\sigma$\,Êerror bars.}
\label{fig_omc}
\end{center}
\end{figure}

The final radial velocities are given in Table~\ref{table_rv} and displayed in Figs.~\ref{fig_omc} 
and~\ref{fig_orb_phas}. The error bars are the quadratic sums of the different error sources 
(photon noise, wavelength calibration and drift, moonlight correction).

\section{Determination of the planetary system parameters}

\subsection{Refined orbit}
\label{sect_refined_orbit}

The radial velocities measurements presented by \jk07\ have a typical accuracy of $\sim160$~\ms.
Those secured with \sophie\ are about five times more accurate, so they allow for a 
refinement of the original parameters of the system. 
We made a Keplerian fit of the first four \sophie\ measurements performed during the transit night 
and those performed afterwards, at other phases, first 
using the orbital period from \jk07. For the refinement of the orbit we did not use most of the data 
secured during the night of January 28th, 2008, in order to remain free from alteration due to
transit anomalies and possible systematic effects due to large-airmass~observations. 

\begin{figure}[h] 
\begin{center}
\vspace{1cm}
\includegraphics[scale=0.445]{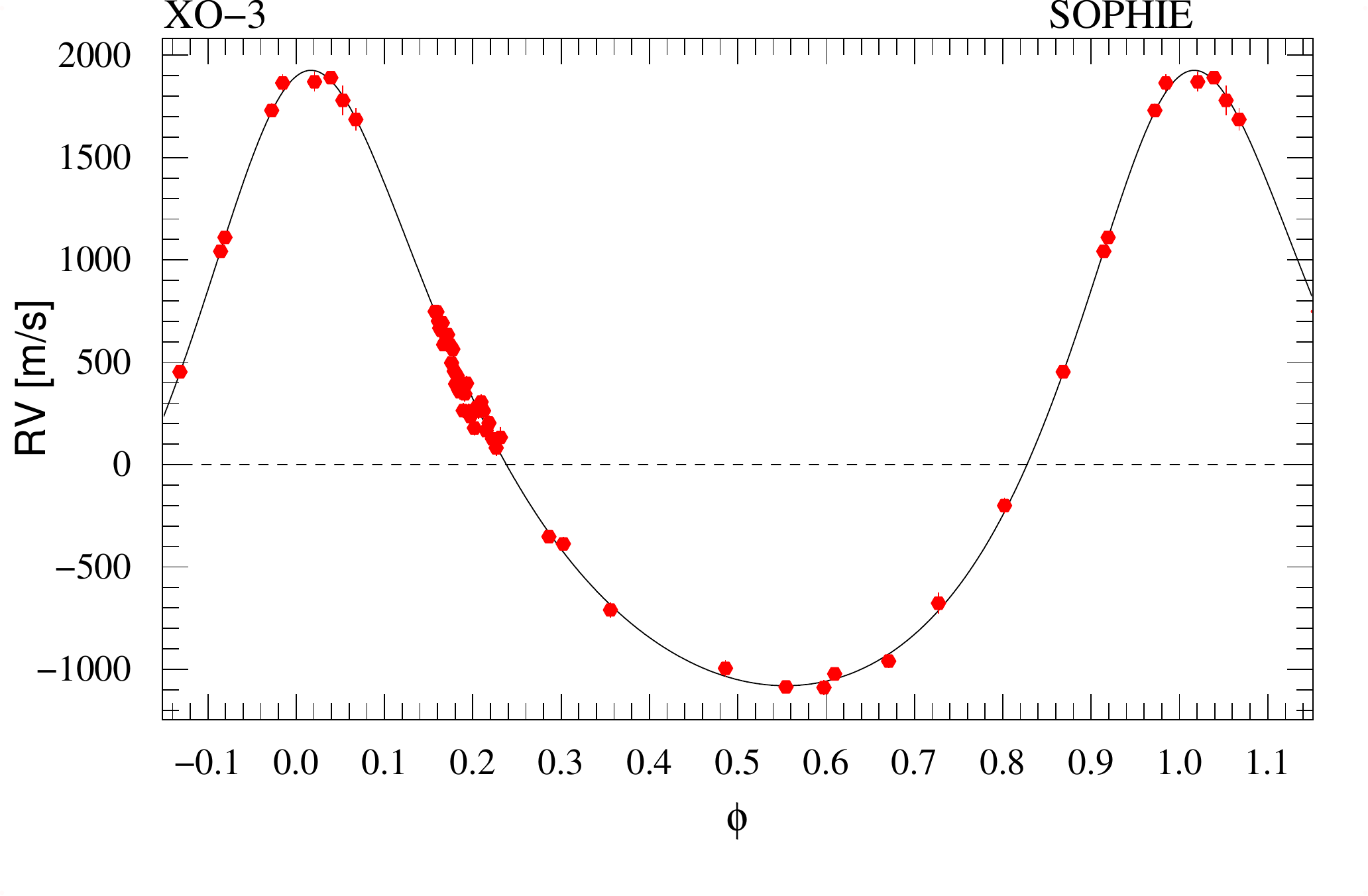}
\caption{Phase-folded radial velocity measurements of XO-3 
(corrected from the velocity $V_r=-12.045\,$\kms)
as a function of the 
orbital phase, and Keplerian fit to the data. Orbital parameters corresponding~to this 
fit are reported in Table~\ref{table_parameters}. For display purpose, all the 
measurements performed during the transit night are plotted here. However,~only the 
first four measurements of the transit night are used for the orbit~fit, together with 19 
measurements secured at other orbital phases~(see~\S~\ref{sect_refined_orbit}).
Figs.~\ref{fig_rm} and \ref{fig_rm_70} display a magnification on the  transit night~measurements.
}
\label{fig_orb_phas}
\end{center}
\end{figure}

The standard deviation of the residuals to the fit is $\sigma(O-C)=29$~\ms, implying a 
\kid\  of 15.3, which is acceptable according the low degrees of freedom, $\nu=18$. 
The 29 \ms\ dispersion of the measurements around the fit is similar 
to the errors on the individual radial velocity measurements; these estimated error bars 
thus are approximatively correct. 
The fits are plotted in  Figs.~\ref{fig_omc} and~\ref{fig_orb_phas};  
the derived orbital parameters are reported in 
Table~\ref{table_parameters}, together with error bars, which were 
computed from  \kid\  variations and Monte~Carlo experiments. They agree 
with the \jk07\ parameters but the error bars are reduced by factors of three to six. 
The largest difference is on the eccentricity, which we found 1.6$\,\sigma$ larger than  
\jk07. The residuals, plotted as a function of time in the bottom panel of 
Fig.~\ref{fig_omc}, do not show any trend that might suggest the presence 
of another companion in the system over two months.

\begin{table}[h]
  \centering 
  \caption{Fitted orbit and planetary parameters for XO-3b.}
  \label{table_parameters}
\begin{tabular}{lcc}
\hline
\hline
Parameters & Values and 1-$\sigma$ error bars & Unit \\
\hline
$V_r$ 				& $-12.045\pm0.006$ 			&   \kms	\\
$P$ 					& $3.19161\pm0.00014$			&   days 	\\
$e$					& $0.287\pm0.005$				\\
$\omega$ 			& $-11.3\pm1.5$				&   $^{\circ}$ \\
$K$					& $1.503\pm0.010$				&   \kms	\\
$T_0$ (periastron)		& $2\,454\,493.944\pm0.009$		&   BJD 	\\
$\sigma(O-C)$			& 29 							&   \ms	\\	
reduced \kid			& 0.85 \\ 
$N$					& 23 \\
$t_c$ (transit)			& $2\,454\,494.549\pm0.014$		&   BJD	\\
$M_\star$				&	$1.3\pm0.2$				&   M$_\odot$	 \\
$R_\star$				&	$1.6 \pm 0.2$				&   R$_\odot$ \\
$M_\textrm{p} \sin i$	 	&	$12.4 \pm 1.9$$^\dagger$	&   M$_\mathrm{Jup}$ \\
$i$					&	$82.5 \pm 1.5$				&   $^{\circ}$\\
$M_\textrm{p}$			&	$12.5 \pm 1.9$$^\dagger$	&   M$_\mathrm{Jup}$\\
$R_\textrm{p}$			&	$1.5 \pm 0.2$				&   R$_\mathrm{Jup}$ \\
$\lambda$			&	$70 \pm 15$				&   $^{\circ}$\\
\hline
$\dagger$: using $M_\star = 1.3\pm0.2$\,M$_\odot$
\end{tabular}
\end{table}

Fitted alone, the 23 \sophie\ measurements have too short time span (60 days) to
measure the period more accurately than \jk07\  from photometric 
observations of twenty transits. A 1.5-year time span is obtained when the \sophie\ 
measurements are fitted together with the radial velocities measured by \jk07\ using the 
telescopes Harlan J. Smith (HJS) and Hobby-Eberly (HET). This longer time span
allows a more accurate period measurement. We obtained $P=3.19168\pm0.00015$~days from the 
fit using the three datasets, in agreement with the photometric one, and with a
similar uncertainty. The final period reported in Table~\ref{table_parameters}
($P=3.19161\pm0.00014$~days) reflect these two measurements and is used for the fits plotted  
in Figs.~\ref{fig_omc} and~\ref{fig_orb_phas}. Adding HJS and HET data does not significantly change 
the other orbital parameters nor their uncertainties. For the global fit using the radial velocities from 
the three instruments, we did not use the last HET measurement, performed during a transit 
(see Sect.~\ref{sect_Conclusion}).

The Keplerian fit of the new \sophie\ radial velocity measurements also improves the transit 
ephemeris, as the photometric transits reported by \jk07\ were secured between December 
2003 and March 2007, one hundred or more XO-3b revolutions before the January 28th, 
2008 transit. The mid-point of this transit predicted from the Keplerian fit of the \sophie\ 
radial velocity measurements is 
$t_c=2\,454\,494.549\pm0.014$ (BJD), i.e. just a few minutes earlier than the prediction from 
\jk07. The uncertainty on this transit mid-point is $\pm20$ minutes (or $\pm0.004$ in orbital phase). 

\begin{figure}[h]
\begin{center}
\vspace{0.3cm}
\includegraphics[scale=0.345]{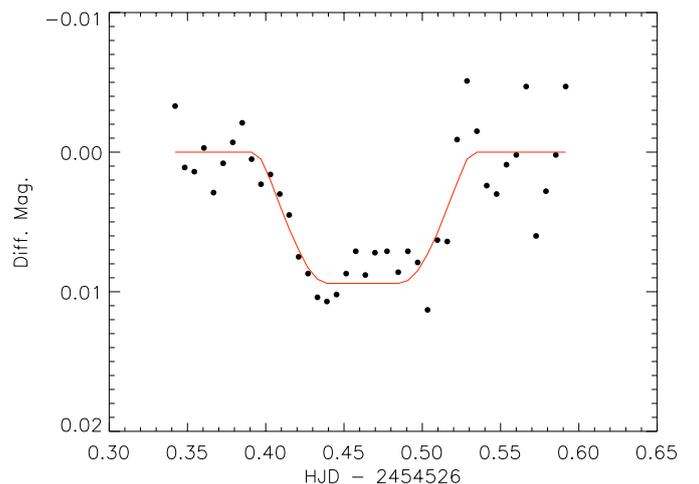}
\caption{Light curve of  XO-3 observed at the Teide Observatory, Tenerife, during 
the transit of February 29th, 2008. The transit fit (solid line) provides 
$t_c=2\,454\,526.4668 \pm 0.0026\equiv 2\,454\,494.5507 \pm 0.0030$~(BJD).}
\label{fig_tran_phot}
\end{center}
\end{figure}

In order to reduce this uncertainty, we observed a recent photometric transit of XO-3b with a 
30-cm telescope at the Teide Observatory, Tenerife, Spain, on February 29th, 2008 
(Fig.~\ref{fig_tran_phot}). Weather conditions were poor and we therefore analyzed 
the transit with a fixed model based on the algorithm of Gim\'enez~(\cite{gimenez06b}). 
The fixed parameters were 
the ratio between the radii of the star and of the planet $k=0.0928$, 
the sum of the projected radii $rr=0.2275$,
the inclination $i=79.3^{\circ}$, 
and the eccentricity $e=0.26$. 
We then scanned different mid-transit times 
and found $t_c=2\,454\,526.4668 \pm 0.0026$ (BJD) from \kid\  variations. This reflects 
photon noise only; fluctuations due to poor weather may introduce additional uncertainties. 
By taking into account for the uncertainty on the orbital period, this translates into 
$t_c = 2\,454\,494.5507 \pm 0.0030 $ (BJD) for the spectroscopic transit that we observed with 
\sophie\ on January 28th, 2008, i.e. ten revolutions earlier. That is just two minutes after the above 
prediction from \sophie\ ephemeris, and the uncertainty on this transit mid-point is 
$\pm4.3$~minutes  (or $\pm0.0009$ in orbital phase).

\subsection{Transit light curve fit revisited}
\label{Transit_light_curve_fit_revisited}

Johns-Krull et al.~(\cite{jk07}) point out that the host star radius 
obtained from the spectroscopic parameters (temperature, gravity, metallicity) combined with 
stellar evolution models, $R_\star \simeq 2.13$~R$_\odot$, is incompatible with 
the value obtained from the shape of the transit light curve, namely 
$R_\star \simeq 1.48$~R$_\odot$. Indeed, a large stellar radius 
implies a large planetary radius (to account for the depth of the transit) and a large inclination 
angle (to account for the duration of the transit), but the time from the first to the second contacts 
(ingress) and third to fourth contacts (egress) predicted for such an inclination are too long when 
compared to the observed transit light curve (see the upper panel of Fig.~9 in Johns-Krull et al.~\cite{jk07}). 
Formal uncertainties on the stellar spectroscopic
parameters and the photometric measurements are insufficient to account for the mismatch.
Since there can be only one value of the real stellar radius, this must be due to systematic 
uncertainties on the spectroscopic parameters, or the parameter derivation from the photometric 
data, or both. We revisit these analyses below, using the photometric data from 
Johns-Krull et al.~(\cite{jk07}) and the parameters of the Keplerian orbit  
obtained in \S\,\ref{sect_refined_orbit} from the \sophie\ radial velocity measurements.

Regarding the spectroscopic parameters, the formal uncertainties stated by Johns-Krull et al.~(\cite{jk07}), 
e.g. 0.06~dex for the gravity $\log g$ or 0.03~dex for the metallicity $Z$, are particularly small. Since these 
are used in combination with stellar evolution models, even if the actual uncertainties on the observations  
are small, systematic uncertainties are known to be present in the models themselves. Also, precise 
gravity measurements are difficult to obtain from stellar spectra. We therefore set a floor level of effective 
uncertainties in the confrontation with stellar evolution  models of 100 $K$ in temperature, of 0.1~dex in 
$\log g$, or 0.1~dex in $Z$  (see e.g. discussion in Santos et al.~\cite{santos04} and Pont \& 
Eyer~\cite{pont04}).

Regarding the photometric data, we estimated the uncertainties including 
systematics effects with ``segmented bootstrap'' analysis  (Jenkins et al.~\cite{jenkins02}; Moutou 
et al.~\cite{moutou04}). According to Pont et al.~(\cite{pont06}), correlated noise usually dominates 
the total parameter error budget for ground-based transit light curves. The segmented bootstrap 
consist of repeating the fit on realizations of the data with individual nights selected at random. 
The photometric follow-up for XO-3 by Johns-Krull et al.~(\cite{jk07}) consists of ten individual
nights. Since the sequencing of the data within each night is preserved, this method provide error 
estimates that takes into account the actual correlated noise in the data.
We find much larger uncertainties on the impact parameter than the photon-noise uncertainties. 
This is corroborated by the discussion in Bakos et al.~(\cite{bakos06}) 
of the case of HD\,189733. With a much deeper transits and a similar number of 
high-precision photometry transits covered from several observatories, they found that the 
determination of the stellar radius from the photometric data produced an error of $\sim15$~\%, 
consistent  with the discussion in Pont et al.~(\cite{pont06}).

\begin{figure}[h]
\begin{center}
\vspace{-0.8cm}
\includegraphics[scale=0.43]{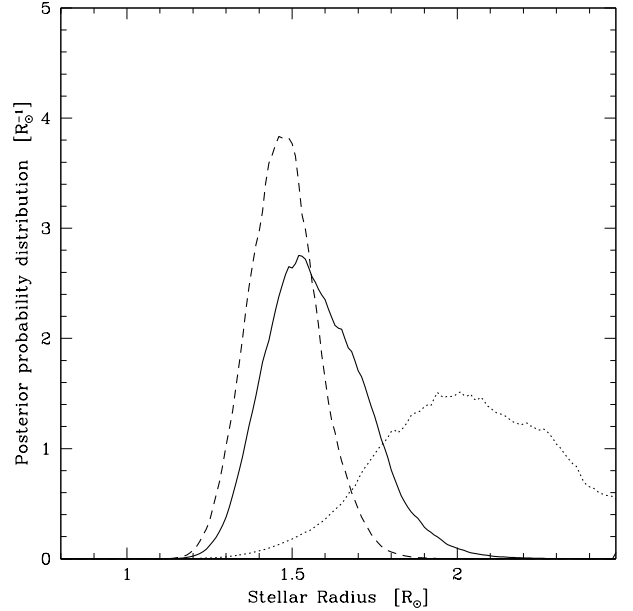}
\vspace{-2cm}
\caption{Posterior probability distribution function for the stellar radius of XO-3 obtained from  
Bayesian approach. 
\textit{Dashed line}: Using only the constraints from the light curves of Johns-Krull et al.~(\cite{jk07}) 
and the parameters of the Keplerian orbit  (\S\,\ref{sect_refined_orbit}). 
\textit{Dotted line}: Using only the constraints from the spectroscopic parameters.
\textit{Solid line}: Using all these constraints together. 
}
\label{fig_pont}
\end{center}
\end{figure}

To estimate the probability distribution of the radius of XO-3 given the available photometric and 
spectroscopic observations,  and the a priori assumption that the star is located near theoretical 
stellar evolution tracks, we use a Bayesian approach. As discussed in Pont \& Eyer~(\cite{pont04}), 
such approach is needed for realistic parameter estimates when the uncertainties are not 
small compared to the total parameter space and the relation between parameters and observable 
quantities are highly non-linear. We thus calculate the posterior probability distribution of the stellar 
radius $R_\star$, using Bayes' theorem with  stellar evolution models and  prior probability distributions  
suitable for a Solar-Neighbourhood magnitude-limited sample, as discussed in Pont~\& 
Eyer~(\cite{pont04}) in the context of the Geneva-Copenhaguen~survey. 

The posterior probability distribution for the stellar radius is calculated according to Bayes' theorem:

\[
{\cal P} (R_\star | P S) = \int {\cal  P} (P |R_\star) {\cal P} (S |R_\star) {\cal P} (R_\star) 
\]
where $R_\star$  is the stellar radius, $P$ the photometric observations and $S$ the spectroscopic 
observations. The first two terms on the right are the likelihood of the photometric 
and spectroscopic observations, $\exp(- 1/2 \chi^2)$, the last term is the a priori distribution of $R_\star$. 
The integral covers the mass, age and metallicity parameters. The stellar 
evolution models provide the function $R_\star=R_\star(M_\star,Z,\mathrm{age})$. For more detailed 
explanations of the method see Pont \& Eyer~(\cite{pont04}).

Fig.~\ref{fig_pont}  displays the posterior probability distribution function for the stellar radius obtained 
from this Bayesian approach given the spectroscopic and photometric data from Johns-Krull et 
al.~(\cite{jk07}), the stellar evolution models from Girardi et al.~(\cite{girardi02}), 
and the orbit parameters determined from the Keplerian fit of 
the \sophie\ radial velocity measurements (\S\,\ref{sect_refined_orbit}).
The probability distribution function for the radius of XO-3 is centered near 
$R_\star \simeq1.5$~R$_\odot$, but extends with non-negligible density from 1.3 to 2.0~R$_\odot$. 
It is well described by $R_\star=1.6 \pm 0.2$~R$_\odot$. The corresponding masses are 
$M_\star = 1.3 \pm 0.2 $~M$_\odot$. This is a quantification of our ``best guess'' from the 
present observational data and prior knowledge about field stars. 
These parameters are reported in~Table~\ref{table_parameters}.

\subsection{Transverse Rossiter-McLaughlin effect?}

\begin{figure}[h]
\begin{center}
\vspace{-3cm}
\includegraphics[scale=0.465]{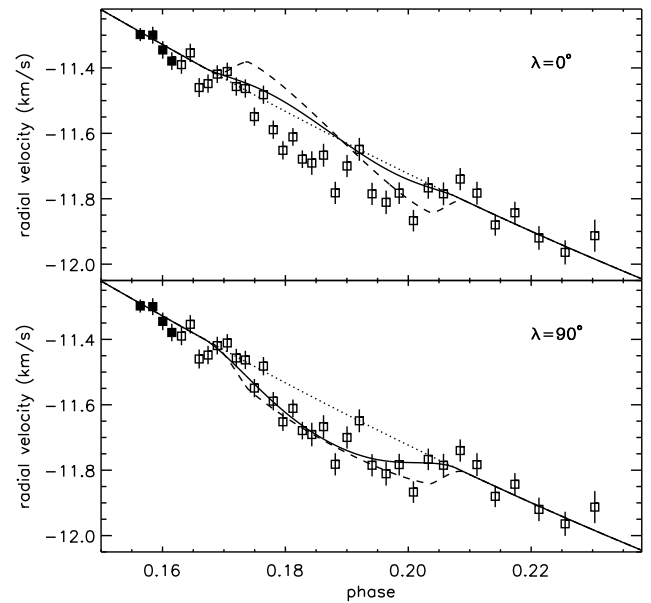}
\vspace{-1.9cm}
\caption{Rossiter-McLaughlin effect models. 
\textit{Top:} $\lambda = 0^{\circ}$ (spin-orbit alignment). \textit{Bottom:} $\lambda = 90^{\circ}$ 
(transverse transit). On both panels, the squares (open and filled) are the \sophie\ radial-velocity 
measurements of XO-3 with 1-$\sigma$\,Êerror bars as a function of the orbital phase. 
Only the 
first four measurements (filled squares) are used for the Keplerian fit (together with 19 
measurements at other orbital phases; see \S~\ref{sect_refined_orbit}).
The dotted line is the Keplerian fit without Rossiter-McLaughlin effect. The two other 
lines show Rossiter-McLaughlin models with $i=78.6^{\circ}$ and $a/R_\star=4.8$ (solid line)
and $i=84.9^{\circ}$ and $a/R_\star=7.2$ (dashed line). The summary of these parameters is in 
Table~\ref{table_rm}. 
}
\label{fig_rm}
\end{center}
\end{figure}

\begin{figure}[h]
\begin{center}
\includegraphics[scale=0.525]{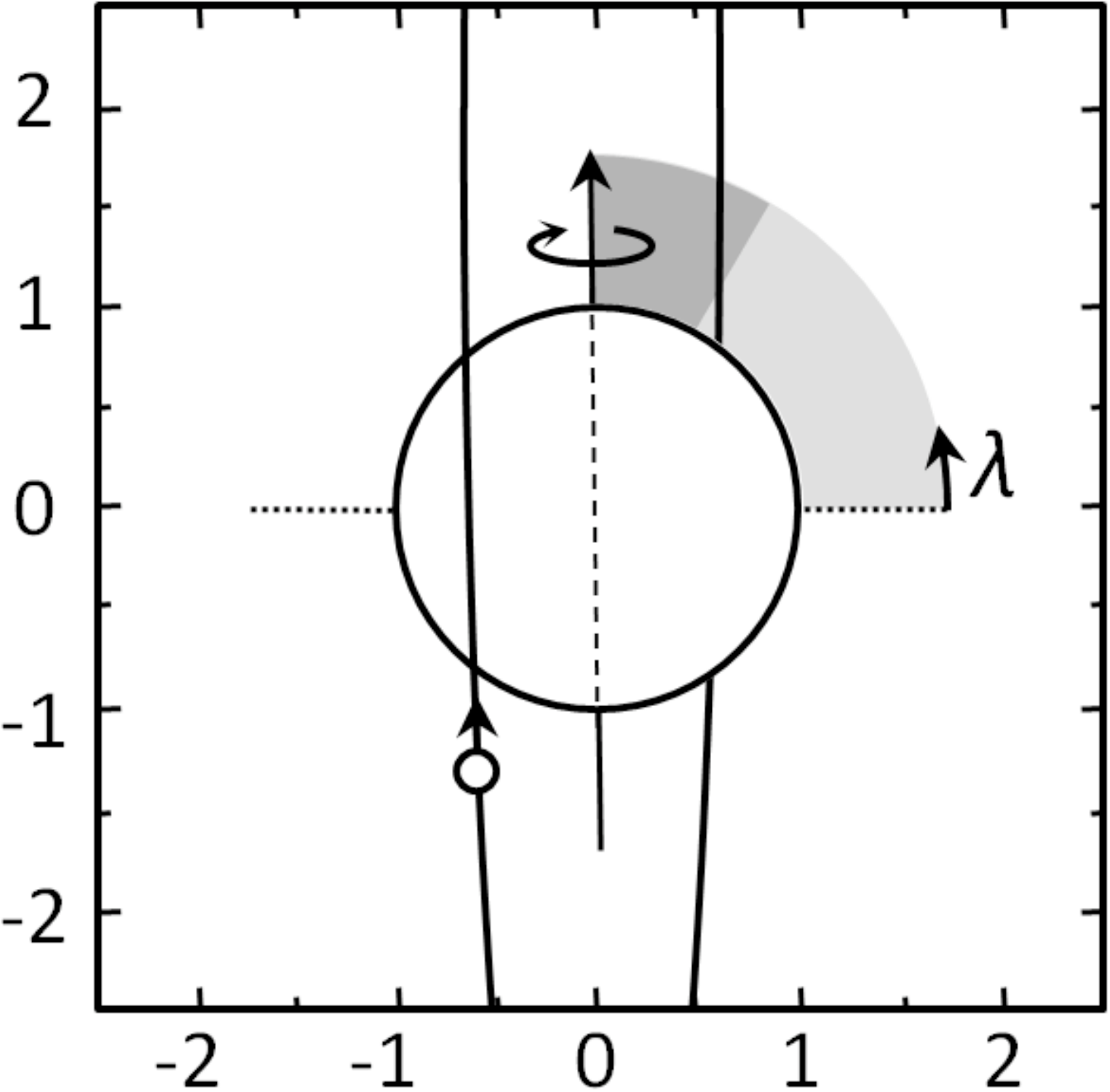}
\caption{Schematic view of the XO-3 system with transverse transit, as seen from the Earth. 
The stellar spin axis is shown, as well as  the planet orbit and 
the $\lambda$ misalignment angle. The scale is in stellar radii.
The limit between soft and strong grey on the $\lambda$-scale represents the favored value 
from our observations ($\lambda =70^{\circ}$, see Sect.~\ref{sect_winn}).}
\label{fig_dessin}
\end{center}
\end{figure}

The radial velocities of XO-3 measured with \sophie\ during the transit of 
January 28th, 2008 are plotted in Fig.~\ref{fig_rm}. Surprisingly, they 
do not show the ordinary anomaly seen in case of prograde transit, i.e. a 
red-shifted radial velocity in the first half of the transit, then blue-shifted in its 
second half. During the full transit of XO-3b, the radial velocity is blue-shifted 
from the Keplerian curve, by about 100~\ms. Such shape is expected 
for a transverse Rossiter-McLaughlin effect, i.e. when the $\lambda$ 
misalignment angle is near $90^{\circ}$ so the the planet crosses the stellar disk
nearly \textit{perpendicularly} to the equator of the star. This is apparently
the case for XO-3b, whose transit seems to only hide some red-shifted 
velocity components, i.e. a part of the star rotating away from the observer.
A schematic view of the XO-3 system with a transverse transit is shown in 
Fig.~\ref{fig_dessin}. 

We overplot in Fig.~\ref{fig_rm} models of Rossiter-McLaughlin effects for XO-3b, for 
$\lambda = 0^{\circ}$ (upper panel) and $\lambda = 90^{\circ}$ (lower panel). 
Following Loeillet et al.~(\cite{loeillet07}) and Bouchy et al.~(\cite{bouchy08}), we used 
the analytical Ohta et al.~(\cite{ohta05}) description of the 
Rossiter-McLaughlin anomaly. We adopted the 
orbital parameters of Table~\ref{table_parameters}, a projected stellar rotation velocity 
$V \sin I$ of 18.5~\kms, and a linear 
limb-darkening coefficient $\epsilon=0.69$ from Claret~(\cite{claret04}), for 
$T_{\mathrm{eff}} = 6250$~K and $\log g=4.0$~dex. The transit was centered 
on the $t_c$ time determined above from the February 29th, 2008 photometric transit. 
To take into account for the large uncertainty in the masses and radii of the 
star and its planet derived from spectroscopic and light curve analyses 
(\S\,\ref{sect_intro} and \S\,\ref{Transit_light_curve_fit_revisited}), we plot the 
models using two extreme sets of parameters 
over the \sophie\ radial velocities in each panel of Fig.~\ref{fig_rm}: 
the solid line is the Rossiter-McLaughlin model with large masses and radii 
  as favored from spectroscopic analyses, and 
the dashed line is the Rossiter-McLaughlin model with smaller masses and radii 
  as favored by the light curve analysis. 
The dotted line is, for comparison, the Keplerian curve without 
Rossiter-McLaughlin~effect.

Table~\ref{table_rm} summaries the parameters used for the different models and the 
quantitative estimations of the quality of the fits. Note that 
the inclination used by Johns-Krull et al.~(\cite{jk07}) for large masses 
and radii, namely $i=79.32^{\circ}$, produces slightly too long transits duration
when used together with our refined orbit. We used $i=78.6^{\circ}$ in that case, 
which remains within the $\pm1.36^{\circ}$ error bar obtained on $i$ by 
Johns-Krull et al.~(\cite{jk07}). 
Models with $\lambda = 0^{\circ}$, or without Rossiter-McLaughlin 
effect detection, produce poor fits, with high \kid\ values and radial velocity 
dispersions of 60 to 75~\ms. This is significantly higher than the expected 
uncertainties on radial velocity measurements, around 33~\ms\ 
(see Table~\ref{table_rv}) and the residuals of the Keplerian fit presented in 
\S\,\ref{sect_refined_orbit}, $\sigma(O-C)=29$~\ms. Thus, our \sophie\  data 
seem to exclude such ordinary~solutions.

The models with $\lambda = 90^{\circ}$ produces lower \kid, with velocity 
dispersions of 42 or 44~\ms. The lower panel of Fig.~\ref{fig_rm} shows that transverse 
transits produce better fits of the data, centered on the expected mid-transit and with the 
adequate duration and depth. One should also note that the \sophie\ measurements 
performed just after the transit (orbital phases from 0.21 to 0.23) are well described by the 
Keplerian orbit model (see Figs.~\ref{fig_orb_phas} and~\ref{fig_rm}). 
We recall that these points were not used to determine the Keplerian 
orbit (\S\,\ref{sect_refined_orbit}); the orbital parameters were determined using only the 
first four measurements of the January 28th, 2008 night
(filled squares in Fig.~\ref{fig_rm}), together with the measurements 
secured on other nights. The good match of the data with the $\lambda = 90^{\circ}$ 
models argues for a transverse transit.
This possible detection is independent of the set of stellar parameters 
adopted in Table~\ref{table_rm}; both produce similar fits with $\lambda = 90^{\circ}$. 
The \kid\ is slightly better in the case of large masses and radii but this does not 
seem to be significant according the noise level.

\begin{table}[h]
\vspace{-1cm}
  \centering 
  \caption{Chosen parameters for the Rossiter-McLaughlin effect models plotted in Fig.~\ref{fig_rm} (see text).}
  \label{table_rm}
\begin{tabular}{cccccccccc}
\hline
\hline
$a/R_\star$ &	$a$ &	$M_\star$	 &	$R_\star$ &	$M_\mathrm{p}$	 &	$R_\mathrm{p}$ &	$i$ &
 	$\lambda$ &	$\sigma$ & \kid\  \\
 & UA &  M$_\odot$ & R$_\odot$ & M$_\mathrm{Jup}$ & R$_\mathrm{Jup}$ & $^{\circ}$ & $^{\circ}$ & \ms &  \\
\hline
4.8 & 	0.048 &	1.4  &	2.1  &	13.7	&	2.0	&	78.6 &	0	&	61 	& 196   \\	
4.8 & 	0.048 &	1.4  &	2.1  &	13.7	&	2.0	&	78.6 &	90	&	42	&   63   \\	
\hline
7.2 &		0.045 &	1.2  &	1.3  &	12.3	&	1.2	&	84.9 &	0	&	74  	&  291   \\	
7.2 &		0.045 &	1.2  &	1.3  &	12.3	&	1.2	&	84.9 &	90	&	44  	&  79	     \\	
\hline
\multicolumn{8}{l}{without transit:} 									&	59 	&  169   \\	
\hline
\end{tabular}
\end{table}

The $\sim40$~\ms\ dispersion of the data from these transverse models remains 
slightly above the computed uncertainties on radial velocity measurements. 
This suggests that some extra uncertainties might be present and not taken into 
account in the error budget. This make us considering this observation 
as a hint of detection for a spin-orbit misalignment.

An explanation for this too large dispersion 
could be the high atmospheric refraction. 
Indeed, as seen in Sect.~\ref{sect_obs}, the end of the transit was observed at large 
airmass. This could introduce biases in the radial velocity measurements that are 
difficult to~quantify.
This agrees 
with the increasing dispersion of the data from these transverse models, which 
is in the range $30-35$~\ms\ in the first half of the transit, then in the range 
$40-45$~\ms\ in the second half.

The larger dispersion might also be partly explained 
as the expected errors are increasing in the second part of the transit 
because of moonlight correction (see \S~\ref{sect_Data_reduction}). 
In addition, it is possible that the planet was 
crossing the stellar disk above a spot; this could cause extra
radial velocity variations (jitter), as the anomaly that is visible near the phase 0.19.
Stellar H and K \ion{Ca}{II} lines do not show core emissions, but they are 
less deep than other F5 stars. This implies $\log R'_{\mathrm{HK}}=-4.6\pm0.2$, 
and we can not exclude XO-3 presents stellar activity, including~spots.

\section{A small radius for XO-3b}
\label{sect_winn}

Shortly after the submission of this paper, photometry of 13 transits 
of XO-3b were released by Winn et al.~(\cite{winn08a}). These new observations 
strongly favors the smaller values for XO-3 and XO-3b radii and masses. The 
parameters reported by Winn et al.~(\cite{winn08a}) agree with the ones presented 
here (Table~\ref{table_parameters}); this strengthen the Bayesian approach we used 
in  \S\ref{Transit_light_curve_fit_revisited}. Timing parameters (as $P$ or $T_0$) from 
Winn et al.~(\cite{winn08a}) are more accurate thanks to their high-quality transit 
photometry, whereas orbital parameters (as $e$, $\omega$ or $K$) are more accurate 
in the present study due to the high-quality radial velocity measurements with~\sophie.

\begin{figure}[h]
\begin{center}
\vspace{-6.6cm}
\includegraphics[scale=0.465]{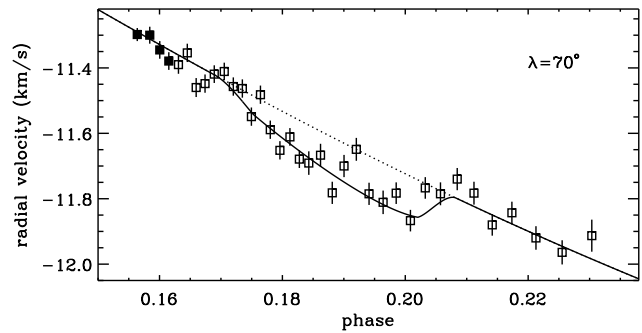}
\vspace{-2.1cm}
\caption{Rossiter-McLaughlin effect models with $\lambda=70^{\circ}$ and the small $R_\mathrm{p}$
value reported by Winn et al.~(\cite{winn08a}). The squares (open and filled) 
are the \sophie\ radial-velocity measurements of XO-3 
with 1-$\sigma$\,Êerror bars as a function of the orbital phase. 
Only the 
first four measurements (filled squares) are used for the Keplerian fit (together with 19 
measurements at other orbital phases; see \S~\ref{sect_refined_orbit}).
The dotted line is the Keplerian fit without Rossiter-McLaughlin effcet. 
The solid and dotted lines are the models with and without Rossiter-McLaughlin effect. }
\label{fig_rm_70}
\end{center}
\end{figure}

\begin{figure}[h]
\begin{center}
\vspace{-3.5cm}
\includegraphics[scale=0.465]{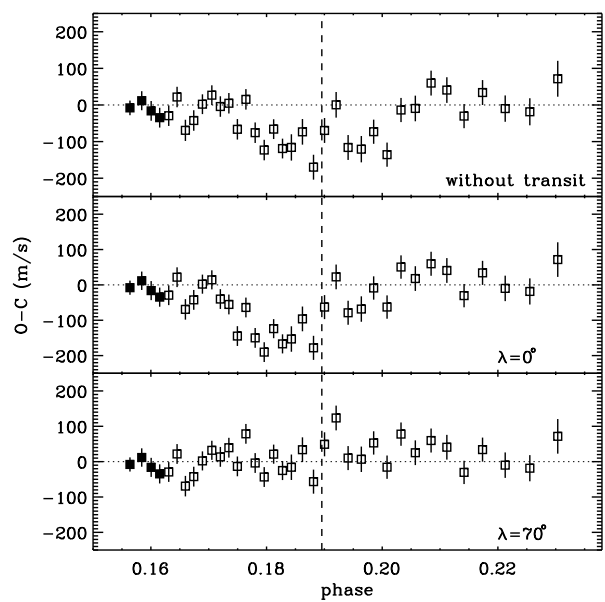}
\vspace{-2.1cm}
\caption{Residuals of the Rossiter-McLaughlin effect fits. 
\textit{Top:} Without transit. 
\textit{Middle:} $\lambda = 0^{\circ}$ (spin-orbit alignment). \textit{Bottom:} $\lambda = 70^{\circ}$. 
The squares (open and filled) are the  \sophie\ radial-velocity measurements of XO-3 with 
1-$\sigma$\,Êerror bars as a function of the orbital phase. 
Only the 
first four measurements (filled squares) are used for the Keplerian fit (together with 19 
measurements at other orbital phases; see \S~\ref{sect_refined_orbit}).
The vertical, dashed line shows the center of the transit.}
\label{fig_rm_70_omc}
\end{center}
\end{figure}

The small $R_\mathrm{p}$ value excludes a grazing transit for XO-3b, and the 
corresponding models plotted in solid lines in Fig.~\ref{fig_rm}; the Rossiter-McLaughlin 
anomaly should thus be large and detectable, with an amplitude 
near the order of magnitude 
$(V \sin I)(R_\mathrm{p}/R_\star)^2 \simeq 150$\,\ms\  (Winn et al.~\cite{winn08a}). 
We fitted the \sophie\ data using the updated parameters from Winn et al.~(\cite{winn08a}), 
in particular
$R_\mathrm{p} = 1.217$~R$_\mathrm{Jup}$, $a/R_\star = 7.07$ and $i=84.20^{\circ}$. 
According \kid\ variations, the range of $\lambda$ compatible with our observations is 
$70^{\circ}\pm15^{\circ}$. The lowest \kid\ is $\sim64$, implying a 42~\ms\ velocity 
dispersions around the model. The best fit with these parameters is plotted in 
Fig.~\ref{fig_rm_70}. The residuals are plotted in Fig.~\ref{fig_rm_70_omc} in three
cases: without transits, with spin-orbit alignment, and with $\lambda =70^{\circ}$. 
Among them, the last case 
is clearly favored by our data when the parameters from 
Winn et al.~(\cite{winn08a}) are adopted.

\section{Conclusion and discussion}
\label{sect_Conclusion}

Table~\ref{table_parameters} summarizes the star, planet and orbit 
parameters of the XO-3 system that we obtained from our analyses. 
The radial velocity measurements that we performed with \sophie\ during a planetary 
transit suggest  that the spin axis of the star XO-3 could be nearly perpendicular to the orbital 
angular momentum  of its planet XO-3b ($\lambda =70^{\circ}\pm15^{\circ}$). 
We note that one Johns-Krull et al.~(\cite{jk07}) HET measurement was obtained 
near a mid-transit of XO-3b. This radial velocity is blue-shifted by ($260\pm194$)~\ms\ 
from the Keplerian curve, in agreement with the possible transverse Rossiter-McLaughlin 
effect we report here, though with a modest significance.

The \sophie\ observation remains noisy,
showing more dispersion around the fit during the transit 
than at other phases. We consider this result as a tentative 
detection of transverse transit rather than a firm detection. Indeed, the end of the 
transit was observed at large airmasses, which could possibly biases the radial velocity 
measurements. Our fits 
favor a transverse transit, 
but one can not totally exclude a systematic error that would
mimic by chance the shape of a transverse transit. This would imply that 
the radial velocities measured during the end of the transit night, at large airmasses, 
would be off by about 100~\ms, i.e. three to four times the expected errors. 
Other spectroscopic transits of XO-3b should thus be observed. They will allow the 
transverse Rossiter-McLaughlin effect to be confirmed or not, and to better quantify its 
parameters, such as the value of the misalignment angle $\lambda$.

Narita et al.~(\cite{narita08}) estimate that the timescale for spin-orbit alignment 
through tidal dissipation is longer than thousand Gyrs. This timescale is 
uncertain, but much longer than the timescale for orbit circularization, which 
itself is longer than the age of XO-3, estimated in the range $2.4-3.1$~Gyrs 
(Johns-Krull et al.~\cite{jk07}); there are thus no obvious reasons to exclude 
an eccentric, transverse system. 
A strong spin-orbit misalignment would favor of formation scenarii 
that invokes planet-planet scattering (Ford \& Rasio~\cite{ford06}) 
or planet-star interaction in a binary system (Takeda et al.~\cite{takeda08})
rather than inward migration due to interaction with the accretion disk.
This suggests in turn that some close-in planets might result from 
gravitational interaction between planets and/or stars.
Chatterjee et al.~(\cite{chatterjee07}) and Nagasawa et al.~(\cite{nagasawa08}) 
have recently shown that scattering with at least three 
large planets can account for hot Jupiters and predicts high spin-orbit inclinations
(see also Malmberg et al.~\cite{malmberg07}). 
On another hand, XO-3b is an object close to the higher end of planetary masses. 
As discussed for instance by Ribas \&  Miralda-Escud\'e~(\cite{ribas07}), there are 
some indications that these objects be low-mass brown dwarfs, formed by gas cloud 
fragmentation rather than core accretion; so that XO-3b may not necessarily
constrain planet formation~scenario.

Finally, pseudo-synchronization might be questioned in the case of the massive 
XO-3b ($M_{\mathrm{XO-3}} \simeq 100 \times M_{\mathrm{XO-3b}}$), 
which moves on an eccentric orbit with a periastron particularly near its host star. 
Tidal frictions might be high enough to tune the stellar rotation velocity close to 
the velocity of its companion on its orbit at the periastron (Zahn~\cite{zahn77}). 
The expected pseudo-synchronized stellar rotation is given by  
$V_\mathrm{rot} = V_\mathrm{p} \times \frac{R_\star}{a(1-e)}$, where 
$V_\mathrm{p} = 2\pi \frac{a}{P}\sqrt{\frac{1+e}{1-e}}$
is the planet velocity at the periastron. For the XO-3 system, this translates into 
$V_\mathrm{rot} \simeq 30 \frac{R_\star}{\mathrm{R}_\odot}$\,\kms\ 
according to the values in Table~\ref{table_parameters}. As the XO-3 
radius is larger than 1.1~R$_\odot$, its rotation velocity $V \sin I = 18.5$~\kms\ 
is clearly smaller than the pseudo-synchronized velocity. However, we note that a 
spin-orbit misalignment would tend to reduce the pseudo-synchronized rotation 
velocity of the star. In that case, the planet approaches at the nearest of its 
star at low stellar latitude, and not above the stellar equator. 
Pseudo-synchronization might thus be possible 
if actually there is a significant spin-orbit misalignment in the XO-3 system.

\begin{acknowledgements}
We thank the technical team at Haute-Provence Observatory for their support with the \sophie\ 
instrument and the 1.93-m OHP~telescope. 
Financial support to  the \sophie\ Consortium 
from the "Programme national de plan\'etologie" (PNP) of CNRS/INSU, 
France, and from the Swiss National Science Foundation (FNSRS) 
are gratefully acknowledged. 
NCS would like to thank the support from 
Funda\c{c}\~ao para a Ci\^encia e a Tecnologia, Portugal, in the form of a grant (references 
POCI/CTE-AST/56453/2004 and PPCDT/CTE-AST/56453/2004), and through program 
Ci\^encia\,2007 (C2007-CAUP-FCT/136/2006).
XB acknowledges support from the Funda\c{c}\~ao para a Ci\^encia e a Tecnologia
(Portugal) in the form of a fellowship (reference
SFRH/BPD/21710/2005) and a program (reference
PTDC/CTE-AST/72685/2006), as well as the Gulbenkian Foundation for
funding through the ``Programa de Est\'{\i}mulo ˆ Investiga\c{c}\~ao''.
AML, AE, and DE acknowledge support from the French
National Research Agency through project grant ANR-NT-05-4\_44463.
MR is funded by the EARA - Marie Curie Early Stage Training fellowship.
   \end{acknowledgements}


\begin{thebibliography}{}

\bibitem[2006]{bakos06} 
Bakos, G. \'A., Knutson, H., Pont, F., et al. 2006, \apj, 650, 1160

\bibitem[2007]{bakos07} 
Bakos, G. \'A., Kov\'acs, G., Torres, G., et al. 2007, \apj, 670, 826

\bibitem[1996]{baranne96} 
Baranne, A., Queloz, D., Mayor, M., et al. 1994, \aaps, 119, 373

\bibitem[2007]{barbieri07} 
Barbieri, M., Alonso, R., Laughlin, G., et al. 2007, \aap, 476, L13

\bibitem[2008]{barbieri08} 
Barbieri, M., et al. 2008, IAU Symposium No. 253 - "Transiting Planets", May 19-23, 2008, Harvard

\bibitem[2008]{barge08} 
Barge, P., Baglin, A., Auvergne, M., et al. 2008, \aap, 482, L17

\bibitem[2005]{bouchy05} 
Bouchy, F., Pont, F., Melo, C., et al. 2005, \aap, 431, 1105

\bibitem[2006]{bouchy06} 
Bouchy, F., and the Sophie team, 2006, in \textit{Tenth Anniversary of 51~Peg-b},
eds. L.~Arnold, F.~Bouchy \& C.~Moutou, 319

\bibitem[2008]{bouchy08} 
Bouchy, F., Queloz, D., Deleuil, M., et al. 2008, \aap, 482, L25

\bibitem[2004]{butler04} 
Butler, R. P., Vogt, S. S., Marcy, Ge. W., et al.
2004, \apj, 617, 580

\bibitem[2007]{chatterjee07} 
Chatterjee, S., Ford, E. B., Rasio, F. A. 2007,  \apj, submitted [arXiv:0703166]

\bibitem[2004]{claret04} 
Claret, A., 2004, \aap, 428, 1001

\bibitem[2007]{cameron07} 
Collier Cameron, A., Bouchy, F., H\'ebrard, G., et al. 2007, \mnras, 375, 951

\bibitem[2007]{fischer07} 
Fischer, D. A., Vogt, S. S., Marcy, G. W., et al. 2007, \apj, 669, 1336

\bibitem[2006]{ford06} 
Ford, E. B., \& Rasio, F. A. 2006, \apj, 638, L45

\bibitem[2007]{gaudi07} 
Gaudi, B. S., \& Winn, J. N. 2007, \apj, 655, 550

\bibitem[2007]{gillon07} 
Gillon, M., Pont, F., Demory, B.-O., Mallmann, F., Mayor, M., Mazeh, T., Queloz, D., 
Shporer, A., Udry, S., Vuissoz, C. 2007, \aap, 472, L13

\bibitem[2006a]{gimenez06a} 
Gim\'enez, A. 2006a, \apj, 650, 408

\bibitem[2006b]{gimenez06b} 
Gim\'enez, A. 2006b, \aap, 450, 1231

\bibitem[2002]{girardi02} 
Girardi, M., Manzato, P., Mezzetti, M., Giuricin, G., Limboz, F. 2002, \apj, 569,~451

\bibitem[2002]{jenkins02} 
Jenkins, J. M., Caldwell, D. A., Borucki, W. J. 2002, \apj, 564, 495

\bibitem[2008]{jk07} 
Johns-Krull, C. M., McCullough, P. R., Burke, C. J., et al. 2008, \apj, 677, 657

\bibitem[2008]{loeillet07} 
Loeillet, B., Shporer, A., Bouchy, F., et al. 2008, \aap, 481, 529

\bibitem[1924]{mclaughlin24} 
McLaughlin, D.B., 1924, \apj, 60, 22

\bibitem[2007]{malmberg07} 
Malmberg, D., Davies, M. B., Chambers, J. E. 2007, \mnras, 377, L1

\bibitem[2004]{moutou04} 
Moutou, C., Pont, F., Bouchy, F., Mayor, M. 2004, \aap, 424, L31

\bibitem[2008]{nagasawa08} 
Nagasawa, M., Ida, S., Bessho, T. 2008,  \apj, 678, 498

\bibitem[2007]{narita07} 
Narita, N., Enya, K., Sato, B., et al. 2007, \pasj, 59, 763

\bibitem[2008]{narita08} 
Narita, N., Sato, B., Ohshima, O., Winn, J. N. 2008, \pasp, 60, L1

\bibitem[2005]{ohta05} 
Ohta, Y., Taruya, A., Suto, Y. 2005, \apj, 622, 1118

\bibitem[2002]{pepe02} 
Pepe, F., Mayor, M., Galland, F., et al. 2002, \aap, 388, 632

\bibitem[2008]{pollacco08} 
Pollacco, D., Skillen, I., Collier Cameron, A., et al. 2007, \mnras, 385, 1576

\bibitem[2004]{pont04} 
Pont, F., \& Eyer, L. 2004, \mnras, 351,  487

\bibitem[2006]{pont06} 
Pont, F., Zucker, S., Queloz, D. 2006, \mnras, 373, 231

\bibitem[2000]{queloz00} 
Queloz, D., Eggenberger, A., Mayor, M., et al. 2000, \aap, 359, L13

\bibitem[2007]{ribas07} 
Ribas, I., \& Miralda-Escud\'e, J. 2007, \aap, 464, 779

\bibitem[1924]{rossiter24} 
Rossiter, R. A., 1924, \apj, 60, 15

\bibitem[2004]{santos04} 
Santos, N. C., Israelia, G., Mayor, M. 2004, \aap, 415, 1153f

\bibitem[2008]{takeda08} 
Takeda, G., Kita, R., Rasio, F. A. 2008, \apj\ in press [arXiv:0802.4088]

\bibitem[2006]{winn06} 
Winn, J. N., Johnson, J. A., Marcy, G. W., et al. 2006, \apj, 653, L69

\bibitem[2007]{winn07} 
Winn, J. N., Holman, M. J., Bakos, G. A., et al. 2007, \apj, 665, L167

\bibitem[2008a]{winn08a} 
Winn, J. N., Holman, M. J., Torres, G., et al.
2008a, \apj, in press [arXiv:0804.4475]

\bibitem[2008b]{winn08b} 
Winn, J. N., Asher Johnson, J., Narita, N., et al. 
2008b, \apj, in press [arXiv:0804.2259]

\bibitem[2007]{wolf07} 
Wolf, A. S., Laughlin, G., Henry, G. W., et al. 2007, \apj, 667, 549
  
\bibitem[1977]{zahn77} 
Zahn, J.-P. 1977, \aap, 57, 383

\end{thebibliography}
\end{document}